\documentclass[twocolumn,showpacs,prl,amsmath,amssymb,superscriptaddress]{revtex4}
\usepackage{color}
\usepackage{verbatim}
\usepackage{hyperref}
\usepackage{epsf}
\usepackage{graphics}
\usepackage{graphicx}
\usepackage{natbib}
\usepackage{dcolumn}
\usepackage{bm}

\begin{document}
\title{Microscopic Route to Nematicity  in Sr$_{3}$Ru$_{2}$O$_{7}$}
\author{Christoph M. Puetter}
\affiliation{Department of Physics, University of Toronto, Toronto,
Ontario, Canada M5S 1A7}
\author{Jeffrey G. Rau}
\affiliation{Department of Physics, University of Toronto, Toronto,
Ontario, Canada M5S 1A7}
\author{Hae-Young Kee}
\email{hykee@physics.utoronto.ca}
\affiliation{Department of Physics, University of Toronto, Toronto,
Ontario, Canada M5S 1A7}
\affiliation{School of Physics, Korea Institute for Advanced Study, Seoul 130-722, Republic of Korea}
%
\begin{abstract}
An anisotropic metallic phase dubbed electronic nematic phase 
bounded by two consecutive metamagnetic transitions 
has been reported in the bilayer ruthenate Sr$_3$Ru$_2$O$_7$. 
It has also been shown that the nematic and 
the accompanying metamagnetic transitions
are driven by an effective momentum-dependent quadrupole-type interaction. 
Here, we study the microscopic origin of such an effective 
interaction.  
To elucidate the mechanism behind the spontaneous 
Fermi surface distortion associated
with the nematic, we identify
a simple tight binding model based
on $t_{2g}$ orbitals, spin-orbit coupling and the rotation of 
RuO$_6$ octahedra as starting point,
consistent with the Fermi surface obtained 
from recent angle-resolved photoemission data.
Within an extended Hubbard model
the nematic state, characterized by an anisotropy between 
the bands near $(\pm \pi,0)$ and $(0,\pm \pi)$, 
then strongly competes with ferromagnetic order
but pre-empts it via a finite nearest neighbor interaction.
We discuss experimental means to confirm our proposal.
\end{abstract}
\maketitle

{\it Introduction} ---
In correlated electron systems, electrons can organize 
themselves in states that are analogous to classical liquid 
crystal phases. \cite{Kivelson98Nature}
The search for such phases in solid-state systems, in particular 
for the quantum version of an anisotropic liquid crystal,
dubbed electronic nematic phase, has been of great interest.
Such a phase spontaneously breaks the point-group symmetry of the 
underlying lattice thus characteristically modifying, e.g., 
transport properties.
Recently, a remarkable anisotropic longitudinal resistivity 
has been reported in the bilayer ruthenate Sr$_{3}$Ru$_{2}$O$_{7}$, 
\cite{Borzi07Science}
where the putative nematic phase is bounded by two consecutive  
metamagnetic transitions. \cite{Grigera04Science}
From a theoretical perspective, important progress 
has been made as well in understanding the phenomenon.
It has been shown that an effective momentum-dependent interaction model
\cite{Oganesyan01PRB,Kee03PRB,Khavkine04PRB} successfully
describes the metamagnetic transitions occurring at the nematic phase 
boundaries, \cite{Kee05PRB} and 
the behavior of the resistivity, when the magnetic field is applied
along the $c$ axis \cite{Doh07PRL} and tilted away from it. 
\cite{Puetter07PRB}

However, the link between the effective interaction and the
microscopic origin is missing.
While attempts to illuminate the mechanism of nematic phase 
formation have been made very recently, \cite{Raghu09PRB,Lee09PRB} 
the focus exclusively lay on the quasi-one-dimensional (1D) Ru $d_{yz}$ 
and $d_{xz}$ orbitals,
identifying the nematic phase as orbital ordering driven 
by an inter-orbital Hubbard interaction.
Yet, there is no a priori reason to exclude the two-dimensional (2D) $d_{xy}$ orbital
from the picture.
On the contrary, angle-resolved  photoemission spectroscopy (ARPES) 
in the isotropic phase 
has revealed a van Hove singularity (vHS) 
near the Fermi level predominantly originating 
from the $d_{xy}$ orbital. \cite{Tamai08PRL}
This is crucial as the isotropic Fermi-surface (FS) structure constrains 
the theoretical starting point for the study of the nematic phase
within a weak-coupling theory 
and the nematic is associated with a distortion of the 
FS topology. \cite{Kee03PRB,Khavkine04PRB}

In this paper, we describe a 
microscopic route to  nematicity in Sr$_{3}$Ru$_{2}$O$_{7}$.
We first model the underlying band structure 
including all three $t_{2\text{g}}$ orbitals
by incorporating unit-cell doubling due to the 
rotation of the RuO$_6$ octahedra 
and spin-orbit (SO) interaction into a single layer approach.
We then show that the nematic phase is associated with a
density imbalance in the $\gamma_{2}$ band
near $(\pm \pi,0)$ and $(0,\pm \pi)$, where the dominant orbital in the 
Bloch function is  the $d_{xy}$ orbital.  
We find that nematic ordering strongly competes with 
ferromagnetic order within a multi-orbital Hubbard model, 
but pre-empts it due to a finite nearest-neighbor interaction.  
Ultimately, we discuss experimental probes to confirm our proposal.

\begin{figure}[t]
  \centering
  \includegraphics*[width=1.0\linewidth, clip]{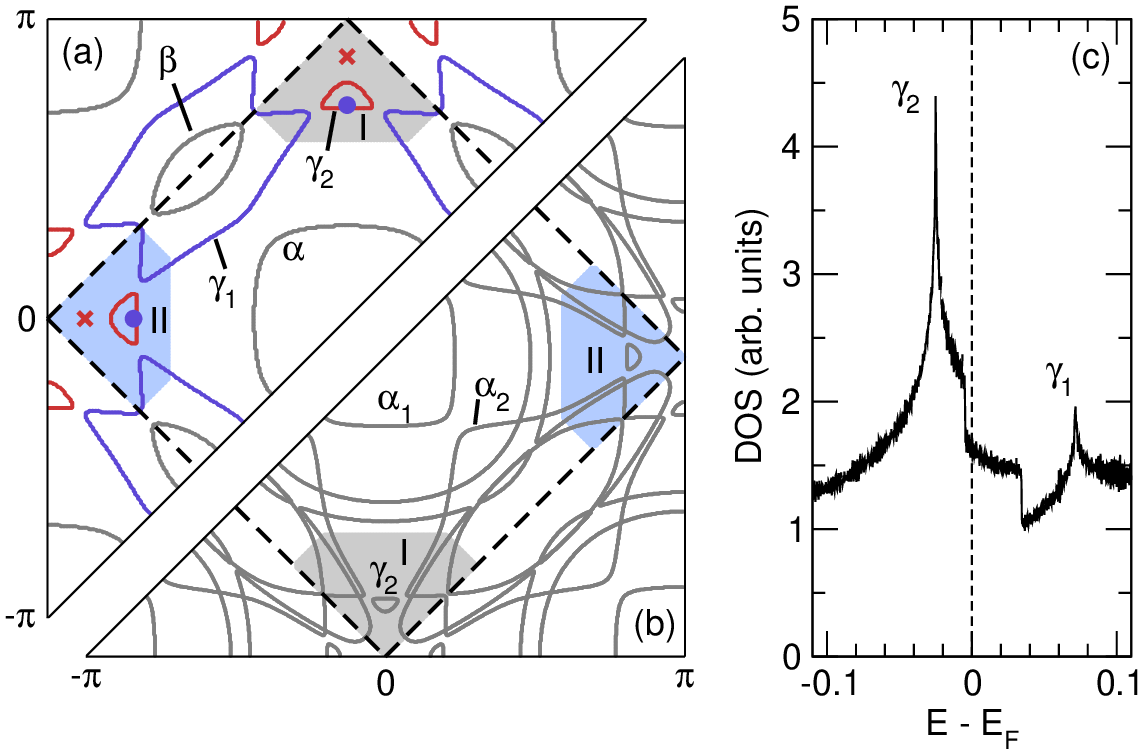}
  \caption{(Color online) (a) Fermi surface of the single layer model. 
    The crosses (dots)
    indicate the saddle points of the $\gamma_{2}$ ($\gamma_{1}$) band.    
    The band structure parameters are
    $t_{1} = 0.5, t_{2} = 0.05, t_{3} = 0.5, t_{4} = 0.1, 
    t_{5} = -0.03, t_{6} = 0.05, \mu = 0.575, {\bar g} = 0.1, 
    \lambda = 0.1375$. 
    (b) Fermi surface of the corresponding bilayer system
    with finite interlayer hopping between quasi-1D orbitals.
    (c) Total DOS near the Fermi level corresponding to (a).
    See main text for details.
    \label{fig:FSNDOS}}
\end{figure}

{\it Fermi-surface topology of Sr$_3$Ru$_2$O$_7$} ---
\label{sec:FS}
To obtain the electronic band structure 
we start from a single layer tight-binding model 
with three $t_{2\text{g}}$ orbitals and SO coupling,
as in the case of the single layer Ruthenate 
Sr$_{2}$RuO$_{4}$. \cite{Shen01PRB,Bergemann00PRL}
However, despite the similarity,
two additional features need to be 
taken into account in the case of Sr$_{3}$Ru$_{2}$O$_{7}$:
(1) a slight rotation of the RuO octahedra 
\cite{Shaked00JSolidStateChem,Fischer09ArXiv}
entailing unit-cell doubling and (2) bilayer coupling.
Here we proceed by including (1) and discuss the role of bilayer coupling later.
The tight-binding band structure $H_{0}$ then takes the form
\begin{eqnarray}
  \label{eq:H0}
  H_{0} &=& \sum_{{\bf k}} \Psi^{\dagger}_{{\bf k} \uparrow} 
  \begin{pmatrix}
    A_{\bf k} & G \\
    G^* & A_{{\bf k}+{\bf Q}} 
  \end{pmatrix}
  \Psi_{{\bf k} \uparrow} 
  + [\text{time reversed}], \nonumber \\
  A_{\bf k} &=&
  \begin{pmatrix}
    \varepsilon^{yz}_{{\bf k}} & \epsilon^{1\text{D}}_{{\bf k}} + \text{i} \lambda & -\lambda \\
    \epsilon^{1\text{D}}_{{\bf k}} - \text{i} \lambda & \varepsilon^{xz}_{{\bf k}} & 
    \text{i} \lambda \\
    -\lambda & - \text{i} \lambda & \varepsilon^{xy}_{{\bf k}}
  \end{pmatrix},
  \; \; 
  G =  {\bar g} {\bf 1}_{3 \times 3},
\end{eqnarray}
where
$\Psi^{\dagger}_{{\bf k} s} = \big( \psi^{yz \dagger}_{{\bf k} s} 
\psi^{xz\dagger}_{{\bf k} s}
\psi^{xy\dagger}_{{\bf k} -s}
\psi^{yz\dagger}_{{\bf k}+{\bf Q} s}
\psi^{xz\dagger}_{{\bf k}+{\bf Q} s}
\psi^{xy\dagger}_{{\bf k}+{\bf Q} -s}
\big)$
consists of fermionic operators creating
an electron with spin projection $s = \uparrow, \downarrow$ 
in one of the three Ru $t_{2\text{g}}$  derived orbitals $\alpha = yz, xz, xy$.
The orbital dispersions are given by
 $\varepsilon^{xz}_{{\bf k}} = - 2 t_{1} \text{cos}(k_{x}) - 2 t_{2} \text{cos}(k_{y}) $, $\varepsilon^{yz}_{{\bf k}} =
  - 2 t_{1} \text{cos}(k_{y}) - 2 t_{2} \text{cos}(k_{x})$, and
  $\varepsilon^{xy}_{{\bf k}} = - 2 t_{3} \big[ \text{cos}(k_{x}) 
  + \text{cos}(k_{y}) \big] - 4 t_{4} \text{cos}(k_{x}) \text{cos}(k_{y})
  - 2 t_{5} \big[ \text{cos}(2 k_{x}) + \text{cos}(2 k_{y}) \big] $,
while $\epsilon^{1\text{D}}_{{\bf k}} = - 4 t_{6} \text{sin}(k_{x}) \text{sin}(k_{y})$
describes the hopping between the two quasi-1D orbitals
and  $2 \lambda \sum_{i} {\bf L}_{i}{\bf S}_{i}$ is the SO interaction
\cite{Ng00EurophysLett}
(all energies in the following are in units of $2t_{1}$).
For simplicity  we have introduced unit-cell doubling via 
an effective 
lattice potential ${\bar g}$ with
modulation vector ${\bf Q} = (\pi, \pi)$.
The resulting Fermi surface and density of states (DOS) near the 
Fermi energy are shown in Figs. \ref{fig:FSNDOS}(a) and 
\ref{fig:FSNDOS}(c),
respectively, where all bands 
are doubly degenerate due to the lack of time-reversal symmetry breaking.

As shown in Fig. \ref{fig:FSNDOS} (a), 
the Fermi surface is composed of four sheets.
The hole-like $\alpha$  
and electron-like $\beta$ sheets are derived from quasi-1D orbitals,  
while the electron-like $\gamma_1$ and hole-like $\gamma_2$ sheets 
arise from quasi-1D and 2D orbital hybridization.
The total DOS in Fig. \ref{fig:FSNDOS} (c) reveals that the 
band structure provides two singularities 
near $E_{\text{F}}$, which originate from
the $\gamma_{1}$ and $\gamma_{2}$ bands. 
Since any instability in a weak-coupling theory is boosted by 
a vHS near the Fermi level, 
these bands are most susceptible to a nematic transition.
Consequently all other bands ($\alpha$ and $\beta$) can be
rendered less important considering the nematic mechanism in Sr$_3$Ru$_2$O$_7$. 
For completeness we also show the effect of bilayer hopping 
in Fig. \ref{fig:FSNDOS} (b).
This FS, consistent with ARPES, exhibits a more complicated structure,
where the $\alpha_{2}$ sheet appears due to bilayer coupling
(but not the $\delta$ sheet, which most likely derives from 
$e_{\text{g}}$ orbitals).
However, since the orbital and topographical
nature of the most singular $\gamma$ bands  near $(\pm \pi,0)$ and 
$(0, \pm \pi)$ remains unaffected by a finite bilayer coupling,
we proceed with the simpler single layer approach.






Indeed, not only are the $\gamma$ bands responsible for the vHS but the corresponding saddle points all lie within
shaded pocket regions I and II in Fig. \ref{fig:FSNDOS} (a) 
(marked by crosses and dots).
In region I (II), the $\gamma_1$ and $\gamma_2$ bands are 
mostly composed of only $d_{yz}$ ($d_{xz }$) and $d_{xy}$ orbitals.
For instance, the flat parts of the $\gamma_2$ pockets 
and of the tips of the $\gamma_1$ pockets 
derive from the quasi-1D orbital, while the curved portions
and the $\gamma_2$ saddle points have a strong $d_{xy}$ character.
Note that the $\gamma_{1}$ and $\gamma_2$ pockets are formed by 
hybridization via SO coupling since even-parity 
perturbations such as ${\bar g}$ do not lift 
the degeneracy between quasi-1D and  2D orbitals due to different parity.

The simple orbital composition of the $\gamma$ bands in regions I and II 
enables us 
to derive effective interactions within the $\gamma$ bands involving only
the highlighted pocket regions.
The singularities in the DOS suggest that either one 
or both $\gamma$ bands are involved in
the formation of a nematic phase
and that ordering can be driven by weak interactions
due to a large DOS near the Fermi level.
One should keep in mind, however, that there is more than 
one possible instability which takes advantage of the vHS.
Below, we study the interactions in the $\gamma_2$ band
in detail and show how different orders compete. 

{\it Microscopic route to nematicity} ---
The bilayer Ruthenate has a metallic ground state. 
Therefore, it is reasonable to assume that long-range interactions 
are well screened, leaving only moderately 
weak on-site and nearest neighbor interactions.
The microscopic Hamiltonian then reads as
\begin{equation}
  \label{eq:MicHamiltonian}
  H = H_{0} + H_{\text{int}}, \nonumber
\end{equation}
where the extended multi-orbital Hubbard interactions are given by
\begin{equation}
  \label{eq:Hint}
  H_{\text{int}} = \! U \! 
  \sum_{i, \alpha} n^{\alpha}_{i \uparrow} n^{\alpha}_{i \downarrow}
  + \tilde{U} \! \mathop{\sum_{i, \alpha \neq \beta}}_{s, s'} 
  n^{\alpha}_{i s} n^{\beta}_{i s'}
  + \! \! \mathop{\sum_{\langle i, j \rangle, \alpha}}_{s, s'} \! \!
  V^{\alpha} n^{\alpha}_{i s} n^{\alpha}_{j s'}
\end{equation}
with
$n^{\alpha}_{i s} = \psi^{\alpha \dagger}_{i s} \psi^{\alpha}_{i s}$
the density operator of the $\alpha$ orbital on Ru site $i$.
Here, $U$, $\tilde U$, and $V^{\alpha}$ represent repulsive on-site 
intra-orbital, on-site inter-orbital, 
and nearest-neighbor intra-orbital interactions, 
respectively.
To derive effective interactions, 
we write $H_{\text{int}}$ in the basis of the Bloch bands, 
which can be accomplished straightforwardly 
as the $\gamma$ bands in regions I and II are predominantly composed
of two orbitals as discussed above.
%
For region I, one thus obtains
\begin{equation}
  \label{eq:pocketapprox}
  \begin{pmatrix}
    a_{{\bf k} \sigma} \\
    c_{{\bf k} \sigma}
  \end{pmatrix}
  =
  \begin{pmatrix}
    \text{cos} \theta_{\bf k} & -\text{sin} \theta_{\bf k} \\
    \text{sin} \theta_{\bf k} & \text{cos} \theta_{\bf k}
  \end{pmatrix}
  \begin{pmatrix}
    \psi^{xy}_{{\bf k}+{\bf Q} s} \\
    \psi^{yz}_{{\bf k}\, -s} 
  \end{pmatrix},
\end{equation}
where  $a_{{\bf k} \sigma}$ and $c_{{\bf k} \sigma}$ represent 
the annihilation of a quasiparticles with momentum
${\bf k}$ and pseudospin $\sigma = \pm$ in the $\gamma_{1}$ 
and $\gamma_{2}$ band, respectively.
The mixing angles are given by 
$\{\text{sin} \theta_{\bf k}, \text{cos} \theta_{\bf k}\}
= 0.5 \Big[1 \pm (\epsilon^{xy}_{{\bf k}+{\bf Q}} - \epsilon^{yz}_{\bf k})
/\sqrt{(\epsilon^{xy}_{{\bf k}+{\bf Q}} - \epsilon^{yz}_{\bf k})^2 
+ |\Gamma^{xy, yz}_{\bf k}|^2}\Big]^{1/2}$, where 
$\Gamma^{xy, yz}_{\bf k}$ denotes the effective orbital
hybridization mainly caused by SO interaction.
Likewise, the $\gamma$ bands can be approximated in region II
by substituting the $d_{yz}$ for the $d_{xz}$ orbital
(or, equivalently, by exchanging $k_{x} \leftrightarrow k_{y}$).

\begin{figure}[t]
  \centering
  \includegraphics*[width=1.0\linewidth, clip]{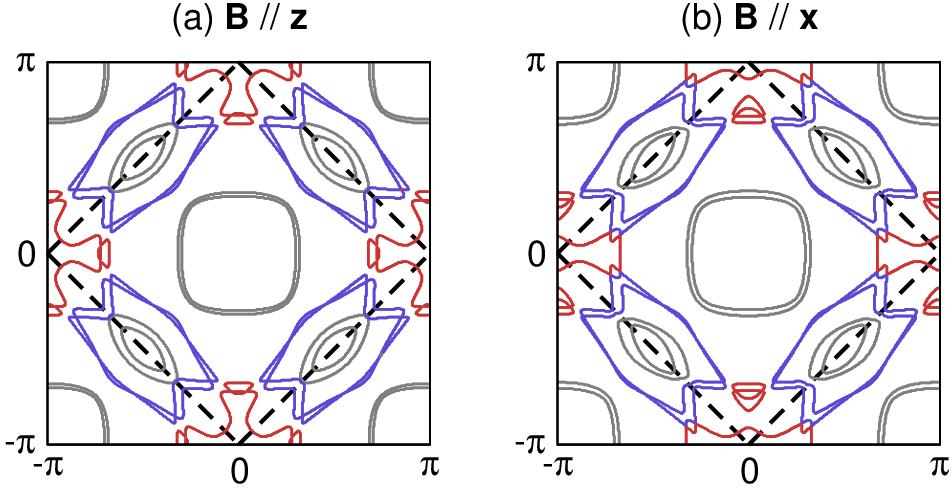}
  \caption{(Color online) Effect of an external magnetic 
    field on the band structure for  
    field orientations (a) parallel to the $z$ axis 
    and (b) parallel to the $x$ axis with $B = 0.05$.
    \label{fig:FSwMagField2}}
\end{figure}

Focusing on the $\gamma_{2}$ band, the effective 
interaction then has the form
\[
 \mathop{\sum_{{\bf k}_{1}, {\bf k}_{2} \; \in \; \text{I}}}
_{ \;\;\;\;\;\; \text{or} \;   \in \; \text{II}}
  U_{1} \; n_{{\bf k}_{1} +} n_{{\bf k}_{2} -} 
+\mathop{\sum_{{\bf k}_{1} \; \in \; \text{I}}}
  _{{\bf k}_{2} \; \in \; \text{II}}
  \big[
  U_{2} \; n_{{\bf k}_{1} \sigma} n_{{\bf k}_{2} \sigma} 
  + U_{3} \; n_{{\bf k}_{1} \sigma} n_{{\bf k}_{2} -\sigma} \big],
\]
where $n_{{\bf k} \sigma} = c^{\dagger}_{{\bf k} \sigma}c_{{\bf k} \sigma}$
and the  interaction strengths are  
\begin{eqnarray}
  \label{eq:EffU}
  U_{1} 
  &=& U \big( \text{sin}^{2}\theta_{{\bf k}_{1}} \text{sin}^{2}\theta_{{\bf k}_{2}}
  + \text{cos}^{2}\theta_{{\bf k}_{1}} \text{cos}^{2}\theta_{{\bf k}_{2}} \big)
  \nonumber \\
  &+&    2 \tilde{U} \text{sin}^{2}\theta_{{\bf k}_{1}} 
  \text{cos}^{2}\theta_{{\bf k}_{2}} 
  + 4 V \text{sin}^{2}\theta_{{\bf k}_{1}} \text{sin}^{2}\theta_{{\bf k}_{2}},
  \nonumber \\
  U_{2}
  &=& \tilde{U} \big(
  \text{sin}^{2}\theta_{{\bf k}_{1}} \text{cos}^{2}\theta_{{\bf k}_{2}} 
  + \text{cos}^{2}\theta_{{\bf k}_{1}} \text{sin}^{2}\theta_{{\bf k}_{2}} \big)
  \nonumber \\
   & + & \tilde{U} \text{cos}^{2}\theta_{{\bf k}_{1}} 
  \text{cos}^{2}\theta_{{\bf k}_{2}}  
  + 8 V \text{sin}^{2}\theta_{{\bf k}_{1}} \text{sin}^{2}\theta_{{\bf k}_{2}},
  \nonumber \\
  U_{3}
  &=& U \text{sin}^{2}\theta_{{\bf k}_{1}} \text{sin}^{2}\theta_{{\bf k}_{2}}
  + \tilde{U} \big( 2 \text{sin}^{2}\theta_{{\bf k}_{1}} 
  \text{cos}^{2}\theta_{{\bf k}_{2}} \nonumber \\
& + &  \text{cos}^{2}\theta_{{\bf k}_{1}} \text{cos}^{2}\theta_{{\bf k}_{2}} \big)
  + 4 V \text{sin}^{2}\theta_{{\bf k}_{1}} \text{sin}^{2}\theta_{{\bf k}_{2}}.
\end{eqnarray}
Here we have included only the largest nearest-neighbor contribution 
$V^{xy} = V$.

Obviously, different instabilities compete.
The dominant one is determined by 
the mixing parameters and the bare interaction strengths.
To investigate the qualitative features of the model we take advantage
of the small size of regions I and II and approximate the form factors 
by their averages 
$s^{2} = \langle \text{sin}^{2} \theta_{\bf k} \rangle = 0.656$
and 
$c^{2} = \langle \text{cos}^{2} \theta_{\bf k} \rangle = 0.344$,
which signifies the dominance of the $d_{xy}$ orbital. \cite{MixParams}
$H_{\text{int}}$ then can be decoupled naturally into
a nematic ($n$), a nematic spin-nematic ($nsn$), and a magnetic ($M$) channel,
\cite{footnote2}
\begin{eqnarray}
  \label{eq:OP}
  \Delta_{\text{n}} &=& \sum_{\sigma} 
  (n^{\text{I}}_{\sigma} - n^{\text{II}}_{\sigma}), \nonumber\\
  \Delta_{\text{nsn}} &=& \sum_{\sigma} 
  \sigma (n^{\text{I}}_{\sigma} - n^{\text{II}}_{\sigma}), \nonumber \\
  \Delta_{\text{M}} &=& \sum_{\sigma} 
  \sigma (n^{\text{I}}_{\sigma} + n^{\text{II}}_{\sigma}),
\end{eqnarray}
with $n^{\text{I/II}}_{\sigma} = N^{-1} \sum_{{\bf k} \; \in \; 
\text{I/II}} n_{{\bf k} \sigma}$
and $N$ the number of ${\bf k}$ points within each region.
All in all, one thus arrives at the following mean-field Hamiltonian 
for the $\gamma_{2}$ band
\begin{eqnarray}
  \label{eq:MFH}
  &H^{\text{MF}}& \! \! = \! \!  \frac{1}{N} \! \!
  \mathop{\sum_{{\bf k} \; \epsilon \; \text{I}, \text{II}}}_{\sigma  = \pm}
    \big( E_{{\bf k} \sigma} - V_{\text{n}} \Delta_{\text{n}} \rho_{\bf k}
  - V_{\text{nsn}} \Delta_{\text{nsn}}  \sigma \rho_{\bf k} \\
  && - V_{\text{M}} \Delta_{\text{M}} \sigma \big)   
c^{\dagger}_{{\bf k} \sigma}  c_{{\bf k} \sigma}
   + V_{\text{n}} \Delta^{2}_{\text{n}}
  + V_{\text{M}} \Delta^{2}_{\text{M}}
  + V_{\text{nsn}} \Delta^{2}_{\text{nsn}}, \nonumber
\end{eqnarray}
where $E_{{\bf k} \sigma}$ denotes the $\gamma_{2}$ quasiparticle bands
and $\rho_{\bf k}$ is set to $+ (-)$ for ${\bf k} \; \in \; \text{I (II)}$. 
The effective interaction parameters are given by
\begin{eqnarray}
  \label{eq:MFIntStrengths}
  V_{\text{n}} &=& (-0.5 U + \tilde{U}) c^4 + \tilde{U} s^2 c^2 
  + 4 V s^4,\nonumber\\
  V_{\text{nsn}} &=& 0.5 U c^4 + \tilde{U} s^2 c^2
  + 4 V s^4, \nonumber\\
  V_{\text{M}} &=& 0.5 U c^4 +  \tilde{U} s^2 c^2 + U s^4.
\label{eq:Veff}
\end{eqnarray}
Note that $V$ affects only the nematic channels. 
While the above analysis is carried out for the $\gamma_2$ band, 
it is also valid for the tips of the $\gamma_1$ band 
inside regions I and II with similar 
effective interaction strengths. 
%

{\it Effect of magnetic field and phase diagram} ---
\label{sec:PD}
The anomalous behavior of the bilayer Ruthenate 
is exposed in the presence of an external magnetic field,
which can be included straightforwardly via Zeeman coupling
$H_{\text{B}} = - {\bf B} \sum_{i} ({\bf L}_{i} + 2 {\bf S}_{i})$.
A field along the $z$ axis, however, only renormalizes the
spin-dependent chemical potential due to a quenching of orbital moments.
The FS therefore changes according to the dominant underlying 
spin character at each ${\bf k}$ point [see Fig. \ref{fig:FSwMagField2} (a)].
Tilting the magnetic field toward the $x$ axis introduces anisotropy 
into the system via SO coupling [Fig \ref{fig:FSwMagField2} (b)].
It also
increases the mixing of the $t_{2\text{g}}$ orbitals in the pocket regions,
where the $\gamma_2$ FS sheet undergoes a 
pronounced topological change
and the tips of the $\gamma_1$ sheet are distinctly distorted.
A tilted field therefore implicates
a low (high) longitudinal conductivity parallel 
(perpendicular) to the in-plane field component (not shown), 
in agreement with the experimental findings in the nematic phase.
\cite{Borzi07Science}

\begin{figure}[t]
  \centering
  \includegraphics*[width=1.0\linewidth, clip]{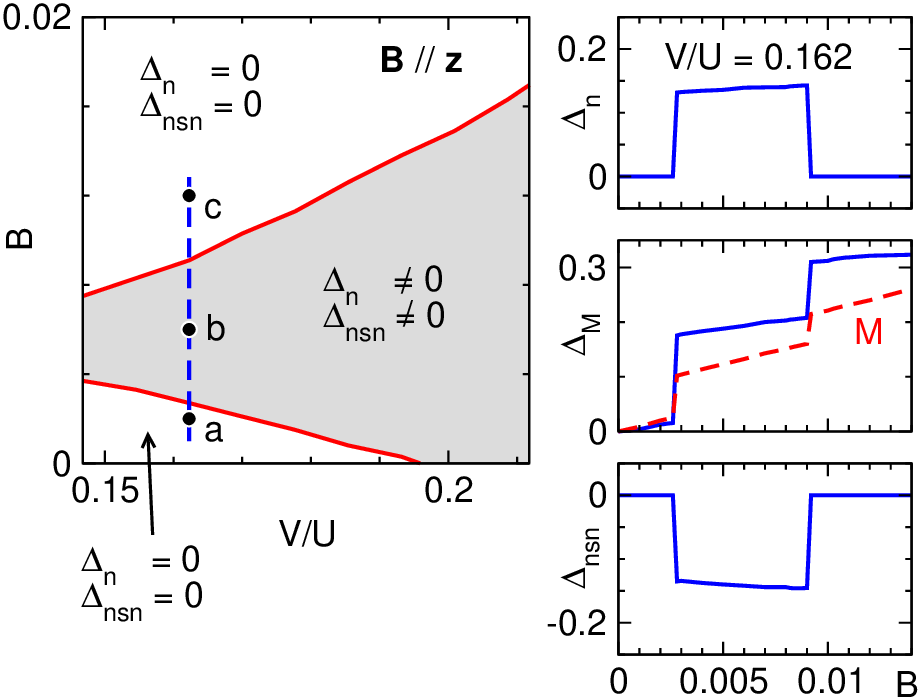}

  \vspace{0.3cm}

  \includegraphics*[width=1.0\linewidth, clip]{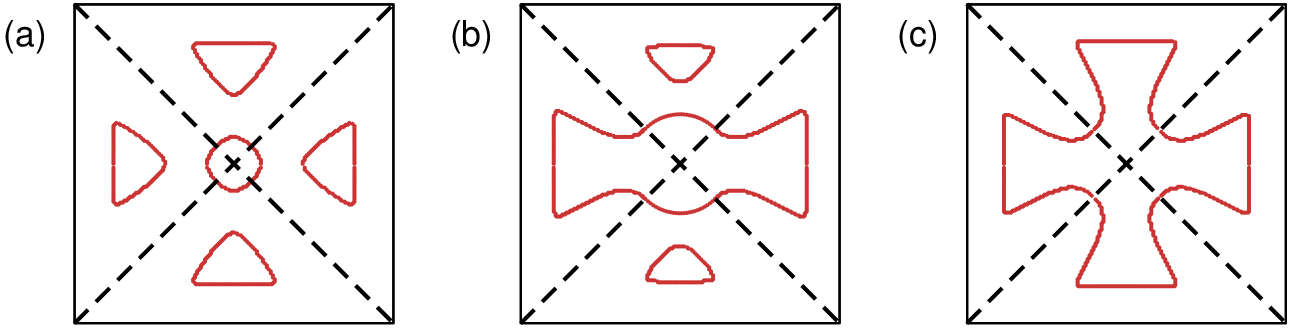}
  \caption{(Color online) Mean field phase diagram.
    The order parameters exhibit dramatic jumps,     
    and the magnetization $M$ is shown as a red dashed line in the
    $\Delta_{\text{M}}$ panel.  
    Panels (a) - (c) display the FS evolution of the nematic
    $\gamma_{2}$ band near $(\pm \pi,0)$/$(0,\pm \pi)$ 
    along the blue dashed line.
    See main text for details.
    \label{fig:MFPhaseDiagram2}}
\end{figure}

In Fig. \ref{fig:MFPhaseDiagram2} we show the mean field phase diagram
as a function of magnetic field along the $z$ axis and $V$ 
for $U = 0.17$, $\tilde{U} = 0.76 U$, 
and the band-structure parameters specified in Fig. \ref{fig:FSNDOS}.
Since the rotation angles and the size of the RuO$_{6}$ octahedra 
vary throughout the ruthenate family, 
\cite{Iliev05PhysicaB,Iwata08JPhysSocJpn,Shaked00JSolidStateChem,Yoshida05PRB}
the nearest neighbor interaction $V$ is the most 
eminent interaction parameter in the present model. 
As expected from Eq. (\ref{eq:Veff}), $V$ promotes nematicity 
with increasing magnitude.
On the other hand, as $V$ decreases, the nematic window shrinks
and the phase boundaries, which are of first order and coincide with
metamagnetic jumps, \cite{Kee05PRB}
merge into a single line of metamagnetic transitions.
Driving the magnetic field strength up at 
small $V/U$, e.g., along the blue dashed line,
tunes one of the $\gamma_{2}$ vHS close enough to the Fermi level
to induce a nematic transition.
This transition may also involve additional bands (such as $\gamma_{1}$) 
due to effective interband interactions.
Note that the effect of bilayer hopping in the presence of an in-plane 
field has been studied previously, \cite{Puetter07PRB}
but is expected to have only a minor influence on nematic ordering.
Note also that in the present context nematic order 
characterizes the difference in the electronic 
or quasiparticle density near $(\pm \pi,0)$ and  $(0,\pm \pi)$. 
The phenomenological order parameter 
$\sum_{{\bf k}} [\cos(k_x) - \cos(k_y)] n_{\bf k}$
used in previous studies
\cite{Kee03PRB,Khavkine04PRB,Kee05PRB,Doh07PRL,Puetter07PRB}
therefore captures the nematic phase remarkably well.

{\it Discussion and summary} ---
We have described a microscopic mechanism for nematicity in the 
bilayer ruthenates.
Our proposal is based on a simple tight-binding approach including 
all three $t_{2\text{g}}$ orbitals, SO interaction, and unit-cell doubling.
Leaving interactions aside, this approach leads to 
a realistic band structure with a FS similar
to recent ARPES data.
Including interactions, 
the most sensitive band is predominantly composed of 
folded $d_{xy}$ and 
unfolded quasi-1D orbitals near $(\pm \pi, 0)$ and $(0, \pm \pi)$, where
it undergoes pronounced anisotropic changes in the nematic phase.
We find that the on-site intra-orbital interaction strongly favors
ferromagnetic order over any nematic order.
On-site inter-orbital interaction on the other hand slightly favors
nematic order over both ferromagnetic and nematic spin-nematic channels
leading to a rather non-trivial and strong competition between
all channels.
However, the nearest neighbor interaction drastically enhances 
charge and spin-nematic susceptibilities,
sufficient to induce a nematic transition in at least one 
of the FS components by applying a moderate magnetic field, 
and thus preempting ferromagnetic order. 
 
Our results imply that nematic order is sensitive to the 
location of vHS and the balance between competing instabilities.
The combination of SO coupling and  
rotation of RuO$_6$ octahedra plays a crucial role in
generating flat bands and the vHS 
near the corners of the reduced Brillouin zone $(\pm \pi, 0)$
and $(0, \pm \pi)$,
where the band structure is most susceptible to change.
Among the members of the Ruthenate family,  
Ca$_{1.8}$Sr$_{0.2}$RuO$_4$ shows a similar metamagnetic transition
\cite{Baier07JLowTempPhys} 
and neutron-scattering pattern \cite{Steffens07PRL} 
as non-ultra-clean Sr$_3$Ru$_2$O$_7$. \cite{Perry01PRL,Capogna03PRB}
The RuO$_6$ octahedra are also rotated, which distinguishes 
this compound from pure Sr$_2$RuO$_4$, in addition to 
the introduction of disorder.
Therefore, it is tempting to argue that there also exists a hidden 
(due to disorder) nematic phase as in the case of
non-ultraclean Sr$_3$Ru$_2$O$_7$. 


Furthermore, a recent experiment probing the entropy landscape
\cite{Rost09Science} has indicated critical behavior outside
the nematic window.
It is possible that criticality arises from 
an uncovered critical point (or weak first-order transition)
outside, yet nearby, the nematic phase space. 
An earlier work \cite{Khavkine04PRB}
has shown that the nematic window shrinks as interactions decrease,
and in fact becomes exponentially small
where critical fluctuations become important. 
However, origin and nature of critical fluctuations 
outside the nematic are beyond the scope of the current 
mean-field study.

It is worthwhile to comment on the difference 
between our proposal and the one in Refs. \cite{Raghu09PRB}
and \cite{Lee09PRB},
where nematic order is associated with orbital ordering between 
the two quasi-1D orbitals, and the anisotropy appears
in the $\alpha_1$ and $\alpha_2$ FS sheets near 
the $\Gamma$ point.  
A possible way to distinguish both proposals is  
scanning tunneling microscopy, \cite{Doh07PRB,Lee09Nature}
which in principle should probe all bands. 
Quantum oscillations would be another way \cite{Mercure09CondMat}
since the $\gamma_{2}$ frequency should drastically 
change between the two isotropic phases indicated in Figs. 
\ref{fig:MFPhaseDiagram2}(a) and \ref{fig:MFPhaseDiagram2}(c),
and, although difficult to observe, should 
split into two when passing through the nematic phase.



\begin{acknowledgments}
We thank Y. Sidis, R. S. Perry,  A. P. Mackenzie, 
A. Green, H. Tagaki, S. Kivelson,
Y. J. Kim, and especially A. Paramekanti for useful discussions.
H.Y.K. acknowledges the University of Tokyo for their hospitality.
This work was supported by NSERC of Canada, Canada Research
Chair, and Canadian Institute for Advanced Research. 
\end{acknowledgments}


\begin{thebibliography}{apssamp}
\bibitem{MixParams} The actual values of the mixing parameters
  near the $\gamma_{2}$ saddle points are larger due to a stronger
  $d_{\text{xy}}$ admixture (thus effectively enhancing 
  $V_{\text{n}}$ and $V_{\text{nsn}}$). 
  However, this only changes quantitative details such as the 
  location of the phase boundary in Fig. \ref{fig:MFPhaseDiagram2}.
\bibitem{footnote2} Considering the FS structure 
  another competing order is a spin density wave (SDW) state, which breaks
  the translational and spin-rotational symmetry. 
  However, it was explicitly shown in Ref. 
  \cite{Raghu09PRB} that SDW order is suppressed by SO coupling. 
  This is consistent with our intuition as the spin is no longer 
  a good quantum number in the presence of SO coupling.
\end{thebibliography}
\end{document}